\documentclass[structabstract]{aa}  
\usepackage{graphicx}
\usepackage{txfonts}
\begin{document}
   \title{High-precision photometry of WASP-12~b transits\thanks{Based on observations collected at the Centro Astron\'omico Hispano Alem\'an (CAHA), operated jointly by the Max-Planck Institut f\"ur Astronomie and the Instituto de Astrofisica de Andalucia (CSIC).}}

   \author{G.~Maciejewski\inst{1,2},
           R.~Errmann\inst{1},
           St.~Raetz\inst{1},
           M.~Seeliger\inst{1},
           I.~Spaleniak\inst{1}, and 
           R.~Neuh\"auser\inst{1},
          }

   \institute{Astrophysikalisches Institut und Universit\"ats-Sternwarte, 
              Schillerg\"asschen 2--3, D--07745 Jena, Germany\\
              \email{gm@astro.uni-jena.de}
         \and
             Toru\'n Centre for Astronomy, N. Copernicus University, 
             Gagarina 11, PL--87100 Toru\'n, Poland
             }
   \authorrunning{G.~Maciejewski et al.}
   \date{Received ...; accepted ...}

 
  \abstract
   {}
   {The transiting extrasolar planet WASP-12~b was found to be one of the most intensely irradiated exoplanets. It is unexpectedly bloated and is losing mass that may accrete into the host star. Our aim was to refine the parameters of this intriguing system and search for signs of transit timing variations.}
   {We gathered high-precision light curves for two transits of WASP-12~b. Assuming various limb-darkening laws, we generated best-fitting models and redetermined parameters of the system. Error estimates were derived by the prayer bead method and Monte Carlo simulations.}
   {System parameters obtained by us are found to agree with previous studies within one sigma. Use of the non-linear limb-darkening laws results in the best-fitting models. With two new mid-transit times, the ephemeris was refined to $\rm{BJD_{TDB}}$ $=$ $(2454508.97682 \pm 0.00020) + (1.09142245 \pm 0.00000033) E$. Interestingly, indications of transit timing variation are detected at the level of 3.4 sigma. This signal can be induced by an additional planet in the system. Simplified numerical simulations shows that a perturber could be a terrestrial-type planet if both planets are in a low-order orbital resonance. However, we emphasise that further observations are needed to confirm variation and to constrain properties of the perturber.}
   {}


   \maketitle
%

\section{Introduction}

WASP-12~b is one of about 100 extrasolar transiting planets known so far. The orbital plane of such a planet is aligned in such a way that the planet passes across the disk of its host star, temporarily blocking a fraction of stellar light. That effect is observed as a flux drop. Combining spectroscopic and photometric data gives a unique opportunity to determine planetary mass and radius, hence density, a key parameter for studying internal structure of an exoplanet.

Hebb et al. (\cite{Hebb}) have found that WASP-12 has a transiting planetary-mass companion. The Jupiter-mass planet orbits the host star with a period of about 1.09 day causing transits with a depth of 14 milli-mag (mmag) and duration of 2.7 hours. The high-accuracy photometric follow-up observations of transits were performed with the robotic 2-m Liverpool Telescope allowing the mid-transit time to be determined with a precision of $\sim$20 s (Hebb et al. \cite{Hebb}). These observations allow the mass, the planet radius, and the orbital semi-major axis to be determined as $M_{\rm{b}}=1.41\pm0.10$ $M_{\rm{J}}$,  $R_{\rm{b}}=1.79\pm0.09$ $R_{\rm{J}}$, and $a=0.0229\pm0.0008$ AU, respectively. The central star is found to be a late-F type dwarf whose age is estimated to be $2 \pm 1$ Gyr. 

The high effective temperature of the central star ($T_{\rm{eff}}=6300^{+300}_{-100}$ K) and short orbital period make WASP-12~b one of the most intensely irradiated exoplanets. This results in a high equilibrium temperature of the planet equal to $2516\pm36$ K when assuming isotropic reradiation and a zero-Bond albedo. Hebb et al. (\cite{Hebb}) note that the internal source of energy may also be needed to explain the unexpectedly large radius of the planet. The orbit of such a close planet is expected to be circularised on short time scales, but surprisingly the orbital eccentricity of WASP-12~b was found to be nonzero ($e=0.049\pm0.015$). Thus, dissipation of tidal energy has been pointed out as a mechanism heating up the planetary interior and bloating the planetary radius (Miller et al. \cite{Miller}). However, further observations showed that WASP-12~b's orbit is highly circular with $e=0.017^{+0.015}_{-0.011}$ (Husnoo et al. \cite{Husnoo}). This result indicates that the eccentricity-driven tidal energy source is negligible.
 
Li et al. (\cite{Li}) predict that the planet may be losing mass at a rate of $\sim$$10^{-7}$ $M_{\rm{J}}$ $\rm{yr}^{-1}$ by exceeding its Roche lobe. The planetary gas is expected to fall towards the host star through Lagrangian point $\rm{L_1}$ and form an optically thin accretion disk. Lai et al. (\cite{Lai}) find by modelling that an accretion stream can have a significant projected area, even comparable to the planet's projected area. The transfer of metals may enhance the apparent stellar metallicity. This effect should be easy to detect because WASP-12 is expected to have a very shallow convective zone. Using spectropolarimetric observations, Fossati et al. (\cite{Fossatia}) find some hints of atmospheric pollution in the photosphere of the host star. Ibgui et al. (\cite{Ibgui}) suggest that the planet is even being disrupted by tidal forces because it may be crossing the Roche limit while passing the periastron. Observing WASP-12~b in near-ultraviolet, Fossati et al. (\cite{Fossatib}) detect many metallic atoms and ions in the planetary exosphere, confirmed to exceed the Roche's lobe. The same conclusion is reached by Croll et al. (\cite{Croll}) on the basis of the analysis of the near-infrared thermal emission of the planet. Moreover, Fossati et al. (\cite{Fossatib}) detect an early ingress and interpret this phenomenon as the presence of previously stripped exosphere matter, which forms a diffuse trail along planet's orbit. The orbital motion could produce a high-opacity bow shock ahead of the planet. Following this idea, Vidotto et al. (\cite{Vidotto}) put an upper constraint on the strength of WASP-12~b's planetary magnetic field, which was found to be $2.4\times10^{-3}$ T.   

L\'opez-Morales et al. (\cite{Lopez}) have detected thermal emission from WASP-12~b by observing its occultations in the $z'$ band and support a non-circular orbit. However, Campo et al. (\cite{Campo}) analysed occultations in the infrared and find that they are centred at phase $\sim$0.5. A model with the orbital precession has been proposed to explain possible variation in timing of occultations. Then, Croll et al. (\cite{Croll}) observed occultations in the near-infrared and combined available photometric and spectroscopic data to argue that the orbital eccentricity of the planet is indistinguishable from 0. No convincing evidence for orbital precession has been found. 

Li et al. (\cite{Li}) note that WASP-12~b's non-zero eccentricity -- if real -- could be excited by an additional super-Earth planet embedded in the circumstellar disk. This hypothesis made WASP-12~b an attractive target to search for transit timing variation (TTV) caused by gravitational interactions with an inner perturber.

\section{Observations and data reduction}

Two transits of WASP-12~b were observed with the 2.2-m telescope at Calar Alto Observatory (Spain) during two runs on 2 and 26 February 2010. During remaining six nights granted to the project, bad weather conditions did not allow us to gather scientific data. The Calar Alto Faint Object Spectrograph (CAFOS) in imaging mode was used as a detector. It was equipped with the SITe CCD matrix ($2048 \times 2048$, 24$\mu$m pixel). A $630 \times 660$ pixel subframe was used during observations to shorten read-out time to about 30 s. The effective field of view was limited to $5\farcm6 \times 5\farcm8$ with the scale of $0.53$ arcsec per pixel, so wide enough to record nearby comparison stars. Including overheads, the observations were carried out with the cadence between 77 and 95 s, depending on the exposure time, which was refined during out-of-transit phase to achieve maximal efficiency. To avoid affecting transit timing, no exposure-time changes were made during the ingress or egress phases. The photometric monitoring was performed in the Johnson $R$-band filter. The telescope was significantly defocused, and stellar profiles exhibited a donut shape. This method was expected to minimise random and flat-fielding errors (Southworth et al. \cite{SouthworthI}, \cite{SouthworthII}, \cite{SouthworthIII}, also Johnson et al. \cite{Johnson}). The stellar flux was spread over a ring of $\sim$24 pixels ($\sim$13 arcsec) in a diameter. This value allowed to avoid contamination of faint neighbour stars, the closest one of which is located $\sim$9 arcsec from the target star. The stellar images were kept exactly at the same position in the detector's matrix during each run thanks to auto-guiding. Our tests showed that in this case dividing by a flat-field frame has no detectable effect on a final light curve. Precise timing was assured by synchronising the computer's clock to Coordinated Universal Time (UTC) by Network Time Protocol software, accurate to better than 0.1 s.   

During the first run, sky conditions were mainly photometric with ephemeral thin clouds before egress. A gap just before the transit's beginning was caused by technical problems. In the second run, observations were gathered in gaps between clouds. Conditions during ingress and egress were photometric and partly cloudy during the flat bottom phase. Observations in both runs were carried out in bright time, and the-second run data were particularly affected by moonlight. The details on observations are presented in Table~\ref{table:1}.

\begin{table}
\caption{The summary of observing runs. } 
\label{table:1}      
\centering                  
\begin{tabular}{c c c c c}      
\hline\hline                
Run\# & Date & $N_{\rm{exp}}$ & $X$ & $T_{\rm{exp}}$ (s) \\ 
\hline                        
   1 & 02 Feb. 2010 & 167 & $1.19\rightarrow1.01\rightarrow1.16$ & 45, 50, 60\\ 
   2 & 26 Feb. 2010 &  69 & $1.01\rightarrow1.40$                & 42, 50, 60\\
\hline                                   
\end{tabular}
\tablefoot{$N_{\rm{exp}}$ -- the number of useful exposures, $X$ -- airmass changes during the run, $T_{\rm{exp}}$ -- exposure times, dates in UT at the beginning of nights.}
\end{table}

The CCD frames were processed using a standard procedure that included subtraction of a dark frame and flat-fielding with dome flats. The magnitudes of WASP-12 and comparison stars were determined with aperture photometry with aperture radii ranging from 8 to 20 pixels. The smallest photometric scatter was found to be at an aperture radius of 12 pixels. This value is also a safety limit to avoid neighbour-star contamination. As a result of iterative tests, two nearby comparison stars, for which airmass and colour differences were the smallest, were used to set the zero level of the photometric scale. The formal photometric error of the individual measurements was found to be in a range between 0.6 and 0.7 mmag. Trends in light curves, caused by differential atmospheric extinction, were modelled using out-of-transit measurements and then removed using the following approximation. For data from 2 February, the second-order polynomial was fitted with the least-square method and then subtracted from the light curve. The out-of-transit data on 26 February were found to be of marginal use, so only a linear trend could be determined and removed.

\section{Results}

The further analysis here was inspired by Southworth (\cite{SouthworthIh}). Individual light curves were modelled with the JKTEBOP code (Southworth et al. \cite{Southwortha, Southworthb}), which is based on the EBOP program (Etzel \cite{Etzel}; Popper \& Etzel \cite{Popper}). Five parameters describing a shape of a light curve were allowed to float during the fitting procedure. We used fractional radii of the host star and planet, defined as $r_{*}=\frac{R_{*}}{a}$ and $r_{\rm{b}}=\frac{R_{\rm{b}}}{a}$, respectively, where $R_{*}$ and $R_{\rm{b}}$ are the absolute radii of the bodies and $a$ is the orbital semi-major axis. In practise, the combinations of these parameters were used, a sum $r_{*}+r_{\rm{b}}$ and ratio $k=r_{*}/r_{\rm{b}}$, because they were found to be only weakly correlated with each other (Southworth \cite{SouthworthIh}). The directly fitted orbital inclination, $i$, allows calculation of the transit parameter $b=\frac{a}{R_{*}}\cos{i}$. The initial values of parameters listed above were taken from Hebb et al. (\cite{Hebb}). To detect any variation in transit timing between observed transits, the mid-transit time was a free parameter whose initial value was calculated according to the ephemeris given by Hebb et al. (\cite{Hebb}). High-quality light curves need to set limb-darkening coefficients (LDCs) as free parameters and test different limb-darkening (LD) laws (Southworth \cite{SouthworthIh}). We therefore considered linear, logarithmic, and square-root LD, for which theoretical LDCs in the Johnson $R$ band were bilinearly interpolated from tables by Van Hamme (\cite{VanHamme}). In a first iteration, we allowed both linear $u$ and non-linear $v$ coefficients to float, but unrealistic values were derived. Therefore, we kept $v$ fixed at theoretical values and allowed only $u$ to vary in final fitting runs. The contribution of $v$ into the error budget was included by perturbing it by $\pm0.1$ around theoretical values and assuming a flat distribution.

The errors of derived parameters were determined in two ways for each combination of a data set and adopted LD law. Firstly, we ran 1000 Monte Carlo (MC) simulations, and a spread range of a given parameter within 68.3\% was taken as its error estimate. Secondly, the prayer-bead method (e.g. D\'esert et al. \cite{Desert}; Winn et al. \cite{Winn}) was used to check whether red noise is present in our data sets. Errors returned by both methods were found to be similar for run\#1, which allowed us to adopt MC errors as the final ones. The run\#2 light curve is of lower quality, and this was reflected in the error analysis. The prayer-bead method gave error values 2-3 times greater than the MC algorithm. This indicates that, besides Poisson noise, there is correlated (red) noise in the light curve. In this case we adopted prayer-bead errors as the final ones. The only exception is the mid-transit time whose MC errors were found to be greater, so they were used as the final ones. This may suggest that the mid-transit time error may still be underestimated. 

\begin{table}
\caption{Parameters of transit light-curve modelling derived for light curves in runs \#1 and \#2.} 
\label{table:2}      
\begin{tabular}{l c c c}      
\hline\hline                
Parameter & Linear & Logarithmic & Square-root\\ 
\hline                        
\multicolumn{4}{c}{Run\#1} \\
$r_{*}+r_{\rm{b}}$      & $0.3850^{+0.0052}_{-0.0048}$    & $0.374^{+0.005}_{-0.005}$       & $0.3710^{+0.0048}_{-0.0051}$   \\ 
$k$                     & $0.11952^{+0.00053}_{-0.00053}$ & $0.11755^{+0.00065}_{-0.00072}$ & $0.1175^{+0.0007}_{-0.0007}$   \\ 
$r_{*}$                 & $0.3439^{+0.0045}_{-0.0042}$    & $0.3342^{+0.0041}_{-0.0042}$    & $0.3320^{+0.0043}_{-0.0045}$   \\ 
$r_b$                   & $0.04110^{+0.00064}_{-0.00064}$ & $0.03928^{+0.00063}_{-0.00068}$ & $0.0390^{+0.0007}_{-0.0007}$   \\ 
$i$ (deg)               & $80.4^{+0.6}_{-0.6}$            & $82.2^{+0.8}_{-0.7}$            & $82.5^{+0.8}_{-0.7}$           \\ 
$b$ ($R_{*}$)           & $0.466^{+0.035}_{-0.034}$       & $0.391^{+0.041}_{-0.038}$       & $0.375^{+0.043}_{-0.039}$      \\ 
$u$                     & $0.35^{+0.03}_{-0.03}$          & $0.571^{+0.054}_{-0.054}$       & $-0.024^{+0.046}_{-0.046}$     \\ 
$T_0$ $(\rm{JD_{UTC}})$ & $0.40144^{+0.00012}_{-0.00011}$ & $0.40146^{+0.00012}_{-0.00010}$ & $0.40144^{+0.00011}_{-0.00010}$\\ 
$rms$ (mmag)            & $0.5896$                        & $0.5840$                        & $0.5839$                       \\ 
$\chi^2_{\rm{red}}$     & $1.1647$                        & $1.1390$                        & $1.1373$                       \\ 
\multicolumn{4}{c}{Run\#2} \\
$r_{*}+r_{\rm{b}}$      & $0.379^{+0.019}_{-0.014}$       & $0.382^{+0.016}_{-0.013}$       & $0.383^{+0.017}_{-0.013}$      \\ 
$k$                     & $0.1192^{+0.0026}_{-0.0021}$    & $0.1188^{+0.0022}_{-0.0016}$    & $0.1190^{+0.0023}_{-0.0015}$   \\ 
$r_{*}$                 & $0.339^{+0.016}_{-0.013}$       & $0.341^{+0.014}_{-0.011}$       & $0.343^{+0.014}_{-0.011}$      \\ 
$r_b$                   & $0.0404^{+0.0027}_{-0.0018}$    & $0.0406^{+0.0023}_{-0.0018}$    & $0.0407^{+0.0024}_{-0.0018}$   \\ 
$i$ (deg)               & $81.9^{+2.4}_{-1.9}$            & $81.8^{+1.8}_{-2.4}$            & $81.6^{+1.8}_{-2.0}$           \\ 
$b$ ($R_{*}$)           & $0.40^{+0.14}_{-0.11}$          & $0.40^{+0.10}_{-0.13}$          & $0.41^{+0.10}_{-0.11}$         \\ 
$u$                     & $0.398^{+0.043}_{-0.035}$       & $0.57^{+0.05}_{-0.08}$          & $-0.017^{+0.064}_{-0.046}$     \\ 
$T_0$ $(\rm{JD_{UTC}})$ & $0.41536^{+0.00013}_{-0.00013}$ & $0.41536^{+0.00013}_{-0.00013}$ & $0.41536^{+0.00013}_{-0.00014}$\\ 
$rms$ (mmag)            & $0.9747$                        & $0.9748$                        & $0.9748$                       \\ 
$\chi^2_{\rm{red}}$     & $2.8718$                        & $2.8664$                        & $2.8660$                       \\ 
\multicolumn{4}{c}{Common parameter} \\
$v$                     & $-$                             & $0.2776$\tablefootmark{a}       & $0.6931$\tablefootmark{a}      \\ 
\hline                                   
\end{tabular}
\tablefoot{$T_0$ is based on UTC and is given as 2455230$+$ and 2455254$+$ for run \#1 and \#2, respectively.}
\tablefoottext{a}{permuted by $\pm0.1$ on a flat distribution}
\end{table}

\begin{figure}
  \centering
  \includegraphics[width=9cm]{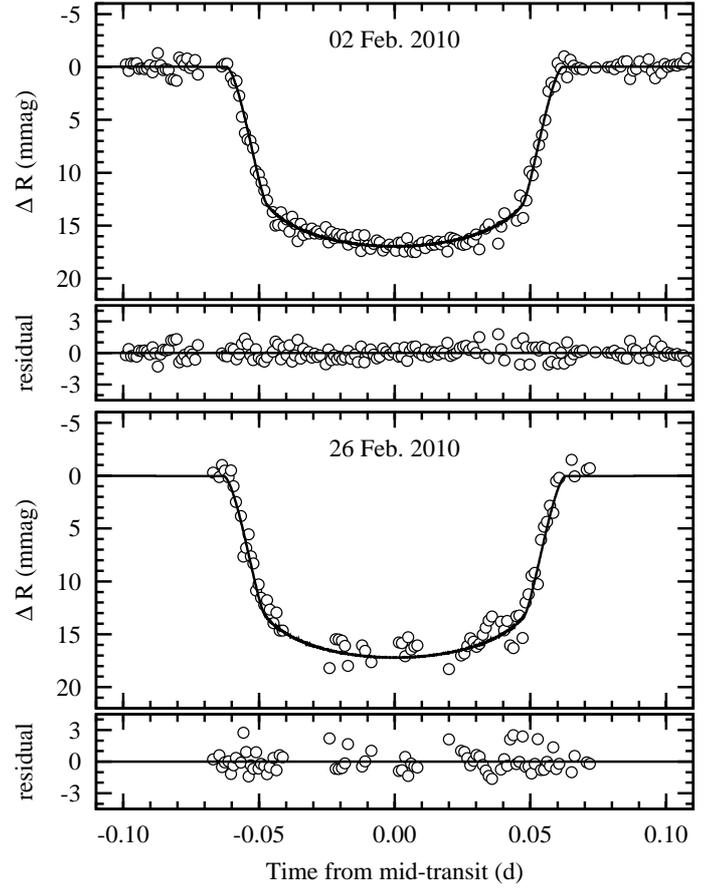}
  \caption{Light curves of two transits of WASP-12~b observed on 2 and 26 Feb. 2010. The best-fitting models, based on the square-root limb darkening law, are plotted with continuous lines. The residuals are shown in the bottom panels.}
  \label{fig-transits}
\end{figure}

Table~\ref{table:2} contains the results of individual fits. For lower quality data, represented by our run\#2 light curve, we found that the goodness of the fit does not depend on the LD law. A significant value of $rms$ (0.97 mmag) clearly indicates that data are affected by non-photometric conditions during the night, so results should be treated with caution. Only the mid-transit time seems to be reliable as non-Poison noise did not affect it. In the case of the run\#1 high-quality data, using the linear LD law results in a slightly poorer fit, while non-linear LD laws give comparable values of $\chi^2_{\rm{red}}$. The same effect was noticed for WASP-4 by Southworth (\cite{SouthworthII}). The light curves acquired in both runs are plotted in Fig.~\ref{fig-transits}, together with best-fitting models and residuals.

\section{Physical properties of WASP-12 system}

Results of light-curve modelling allowed us to calculate planetary, stellar, and geometrical parameters. We used results from the best-fitting model based on the run\#1 data and square-root LD law. The value of $R_{\rm{b}}$ was calculated assuming $a=0.0229\pm0.0008$ AU (Hebb et al. \cite{Hebb}). The planetary mean density $\rho_{\rm{b}}$ uses WASP-12~b's mass of $1.41\pm0.10$ $M_{\rm{J}}$, known from radial velocity measurements. The surface gravitational acceleration, $g_{\rm{b}}$, was calculated according to a formula (Southworth et al. \cite{Southworth2007}): 
\begin{equation}
     g_{\rm{b}} = \frac{2\pi}{P_{\rm{b}}}\frac{\sqrt{1-e^2}}{r^{2}_{\rm{b}} \sin i} K_{*}\, , \; 
\end{equation}
where $P_{\rm{b}}$ is the orbital period of the planet (see Sect.~5), and $K_{*}$ is the stellar velocity amplitude taken from Hebb et al. (\cite{Hebb}). The orbital eccentricity was assumed to be 0. 

The equilibrium temperature, $T_{\rm{eq}}$, was derived by assuming the effective temperature of the host star $T_{\rm{eff}}=6300^{+200}_{-100}$ K (Hebb et al. \cite{Hebb}) and using the relation (Southworth \cite{SouthworthIIIh}): 
\begin{equation}
  T_{\rm{eq}} = T_{\rm{eff}} \left(\frac{1-A}{4F}\right)^{0.25} \left( \frac{r_{*}}{2} \right)^{0.5}\, , \; 
\end{equation}
where $A$ is the Bond albedo and $F$ the heat redistribution factor. This formula may be simplified by assuming relation $A=1-4F$, and the modified equilibrium temperature $T'_{\rm{eq}}$ was derived (Southworth \cite{SouthworthIIIh}). 

We calculated the Safronov number $\Theta$, which is proportional to the ratio of the escape velocity from the planet and velocity of its orbital motion. Southworth (\cite{SouthworthIIIh}) gives formula
\begin{equation}
  \Theta = \frac{M_{\rm{b}}}{M_{*}} r_{\rm{b}}^{-1}\, , \; 
\end{equation}
where the the value of $1.35\pm0.14$ $M_{\odot}$ was taken as the star's mass $M_{*}$ (Hebb et al. \cite{Hebb}). 

The shape of the planet was found to be far from an ideal sphere. The oblateness of the planet, directly returned by the JKTEBOP code and defined as
\begin{equation}
  f = 1 - \frac{r_1}{r_2}\, , \; 
\end{equation}
where $r_1$ and $r_2$ are the polar and equatorial radii, respectively, was found to be 0.083. This locates WASP-12~b between Jupiter and Saturn, whose oblateness values are 0.065 and 0.098, respectively. Results of our calculations are collected in Table~\ref{table:3} where the literature values are also given. For almost all parameters we obtained excellent agreement within $1 \sigma$ error bars with determinations reported by Hebb et al. (\cite{Hebb}). 

\begin{table}
\caption{Physical properties of the WASP-12 system derived from light-curve modelling (see text for explanations).} 
\label{table:3}      
\centering                  
\begin{tabular}{l c c}      
\hline\hline                
Parameter & This work & Hebb et al. (\cite{Hebb})\\ 
\hline                        
\multicolumn{3}{c}{Planetary properties} \\
$R_{\rm{b}}$ $(R_{\rm{J}})$       & $1.9\pm0.1$               & $1.79\pm0.09$          \\ 
$\rho_{\rm{b}}$ $(\rho_{\rm{J}})$ & $0.22\pm0.04$             & $0.24^{+0.03}_{-0.02}$ \\ 
$g_{\rm{b}}$ (m s$^{-2}$)           & $10.3\pm0.6$              & $9.8\pm0.7$            \\ 
$T'_{\rm{eq}}$ (K)                  & $2570^{+100}_{-60}$       & $2516\pm36$            \\ 
$\Theta$                            & $0.026\pm0.005$           & $-$                    \\ 
$f$                                 & $0.083$                   & $-$                    \\ 
\multicolumn{3}{c}{Geometrical parameters} \\
$i$ (deg)                           & $82.5^{+0.8}_{-0.7}$      & $83.1^{+1.4}_{-1.1}$   \\
$b$ $(R_{*})$                       & $0.375^{+0.042}_{-0.039}$ & $0.36^{+0.05}_{-0.06}$ \\ 
$(R_{\rm{b}}/R_{*})^2$              & $0.01380\pm0.00016$       & $0.0138\pm0.0002$      \\ 
\multicolumn{3}{c}{Stellar properties} \\
$R_{*}$ $(R_{\odot})$               & $1.63\pm0.08$             & $1.57\pm0.07$          \\ 
$\rho_{*}$ $(\rho_{\odot})$         & $4.14\pm0.09$             & $4.17\pm0.03$          \\ 
$\log g_{*}$ (cgs)                  & $0.31\pm0.04$             & $0.35\pm0.03$          \\ 
\hline                                   
\end{tabular}
\end{table}

\section{Transit timing}

\begin{figure}
  \centering
  \includegraphics[width=9cm]{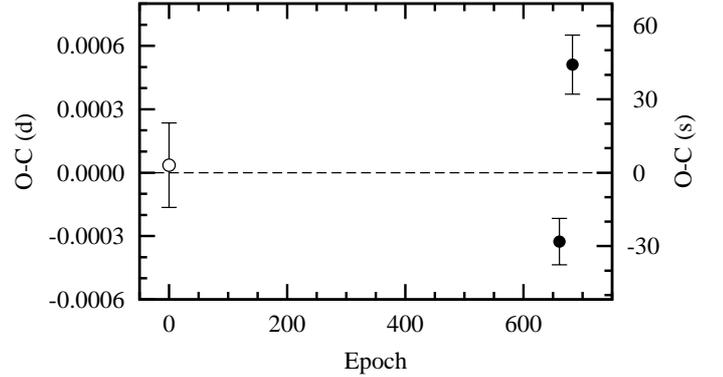}
  \caption{The observation minus calculation ($O-C$) diagram for WASP-12~b generated for a new linear ephemeris. The open symbol comes from Hebb et al. (\cite{Hebb}), and the filled symbols denote mid-transit times reported in this paper.}
  \label{fig-oc}
\end{figure}

The mid-transit times were transformed from JD based on UTC into BJD based on Barycentric Dynamical Time (TDB) using the on-line converter\footnote{http://astroutils.astronomy.ohio-state.edu/time/utc2bjd.html} by Eastman et al. (\cite{Eastman}). Having a long time span of observations, we redetermined an improved ephemeris. As a result of fitting a linear function of the epoch and orbital period $P_{\rm{b}}$, we obtained
\[
	T_{0}  = 2454508.97682 \pm 0.00020 \,  \; \rm{BJD_{TDB}} 
\]
\[
	P_{\rm{b}}      = 1.09142245 \pm 0.00000033 \,  \; \rm{d}. 
\]
Individual mid-transit errors were taken as weights. The observation minus calculation ($O-C$) diagram, plotted in Fig.~\ref{fig-oc}, clearly shows a deviation in our mid-transit times from the new linear ephemeris. The amplitude of the possible TTV of WASP-12~b was found to be 73 s between 22 epochs at the level of $3.4 \sigma$. To check whether keeping all parameters free for run\#2 data may affect the mid-transit time, we repeated modelling run\#2 data with all parameters except the mid-transit time kept fixed. Their values were adopted from the run\#1 analysis. The derived mid-transit time agreed with the full-modelling value well within $1 \sigma$. As the additional test, both light curves were reanalysed with a model-fitting algorithm available via the Exoplanet Transit Database\footnote{http://var2.astro.cz/ETD/} (Poddan\'y et al. \cite{poddanyetal10}) to obtain mid-transit times with an alternative method. The procedure employs the \textsc{occultsmall} routine of Mandel \& Agol (\cite{mandelalgol02}) and the Levenberg--Marquardt non-linear least-square fitting algorithm, which also provides uncertainties. The procedure assumes the linear LD law with the fixed theoretical LDC whose value was taken from tables by Van Hamme (\cite{VanHamme}). We derived $T_{0}$ equal to $2455230.40143\pm0.00012$ and $2455254.41533\pm0.00020$ $\rm{JD_{\rm{UTC}}}$ for runs\#1 and \#2, respectively. Mid-transit times obtained with both codes turned out to be consistent with each other within $1 \sigma$, which strengthens the reliability of the possible TTV detection. The transit-timing results are summarised in Table~\ref{table:4}. 

\begin{table}
\caption{Results of transit timing.} 
\label{table:4}      
\centering                  
\begin{tabular}{c c c c c}      
\hline\hline                
Run\# & $T_{0}$ $(\rm{BJD_{\rm{TDB}}})$ & $T_{0}$ error (d) & Epoch & $O-C$ (d) \\ 
      & $T_{0}$ $(\rm{JD_{\rm{UTC}}})$  &               &       &  \\ 
\hline                        
   1 & $2455230.40673$ & $\pm0.00011$ & 661 & $-0.00033$\\ 
     & $2455230.40144$ &              &     &  \\ 
   2 & $2455254.41887$ & $\pm0.00014$ & 683 & $+0.00051$\\
     & $2455254.41536$ &              &     &  \\
\hline                                   
\end{tabular}
\tablefoot{The mid-transit times, $T_{0}$, are given as both BJD (based on Barycentric Dynamical Time, TDB) and JD (based on Coordinated Universal Time, UTC).}
\end{table}

\begin{figure}
  \centering
  \includegraphics[width=9cm]{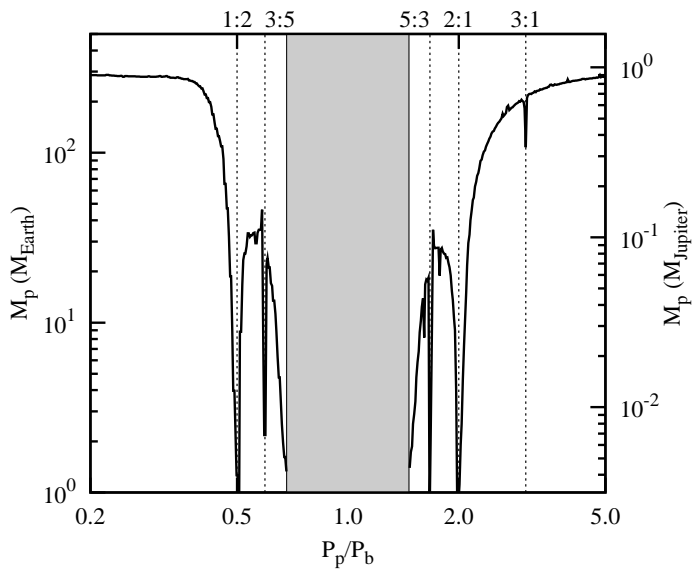}
  \caption{The mass of a perturbing planet that could generate a TTV signal with an observed amplitude of 73 s, as a function of ratio of orbital periods of transiting planet, $P_{\rm{b}}$, and the perturber, $P_{\rm{p}}$. Vertical dashed lines indicate MMRs. Orbits in the range of $P_{\rm{p}}/P_{\rm{b}}$ between $0.69$ and $1.45$, marked by a grey area, were found to be unstable. The case of $P_{\rm{p}}/P_{\rm{b}}=1$ was found to be stable but was skipped in the plot.}
  \label{fig-ttvlim}
\end{figure}

The short time scale of possible TTV signal suggests that it may be caused by an additional low-mass planet perturbing the orbital motion of WASP-12~b (Miralda--Escud\'e \cite{miralda02}; Schneider \cite{Schneider04}; Holman \& Murray \cite{holmanmurray05}; Agol et al. \cite{agoletal05}; Steffen et al. \cite{steffenetal07}). Because of the few points in the $O-C$ diagram, it is impossible to characterise the time scale of the TTV and detect some periodicity. However, by knowing the amplitude of the TTV, which may be treated as a minimal value, one can determine the minimal mass of the additional planet in a wide range of orbital configurations. We generated 745 synthetic $O-C$ diagrams for orbits from the inner 1:5 mean motion resonance (MMR) to the outer 5:1 one. For simplicity we assumed that the planetary orbits are coplanar and initially circular. We used the \textsc{Mercury} package (Chambers \cite{chambers99}) and the Bulirsch--Stoer algorithm to integrate the equations of motion for this three-body problem. Computations covered 700 periods of the transiting planet, i.e. the time span of observations. Figure~\ref{fig-ttvlim} illustrates the results of our simulations. The observed amplitude of the TTV could be caused by a body as small as an Earth-mass planet if it was in one of several low-order MMRs. For instance, such a planet would induce stellar radial-velocity variations with the semi-amplitude of 0.6 and 0.4 m s$^{-1}$ in the inner and outer 2:1 resonances, respectively, which in practice are undetectable with modern state-of-the-art instruments. This finding makes WASP-12~b an attractive target for the TTV follow-ups. 

\section{Concluding remarks}

Our high-precision photometry gathered for two transits of WASP-12~b in February 2010 allowed us to redetermine parameters of this extra-solar planetary system. They were found to agree with the values reported by Hebb et al. (\cite{Hebb}). Two new mid-transit times were used to derive improved ephemeris. We found that the observed transit timing suggests there is a short-time variation at the level of $3.4 \sigma$. Such a TTV signal could be caused by an additional terrestrial-type planet if both planets are close to orbital resonances. Further high-precision observations of WASP-12~b's transits are needed to verify the existence of the TTV signal and put constraints on the second planet's parameters.    
 
\begin{acknowledgements}
The authors are grateful to the staff of the Calar Alto Astronomical Observatory for their support during observing runs. GM, SR, and IS acknowledge support from the EU in the FP6 MC ToK project MTKD-CT-2006-042514. GM acknowledges the financial support from the Polish Ministry of Science and Higher Education through the Iuventus Plus grant IP2010 023070. RN, IS, and RE would like to thank the German national science foundation Deutsche Forschungsgemeinschaft (DFG) for support in grants NE 515/30-1, 32-1, and 34-1. GM, SR, and RN acknowledge support from the DAAD PPP--MNiSW project 50724260--2010/2011 \textit{Eclipsing binaries in young clusters and planet transit time variations}. The authors are grateful to the anonymous referee for remarks improving the manuscript.
\end{acknowledgements}

\end{document}